\documentclass[a4paper,11pt]{article}
\pdfoutput=1 
\usepackage{jcappub} 
\usepackage{lineno}
\usepackage{caption}
\usepackage[dvipsnames]{xcolor}
\usepackage{aasmacros}              
\usepackage{comment}
\usepackage{array}
\usepackage{subcaption}
\usepackage{multirow}
\usepackage{graphicx}
\graphicspath{{graphics/}}
\usepackage{amsmath}
\usepackage{amssymb}
\usepackage{rotating} 
\usepackage{xspace}
\usepackage{xcolor}
\usepackage{booktabs}
\usepackage[normalem]{ulem}
\usepackage{adjustbox}
\usepackage{lipsum}  
\newcommand{\Msun}{{\rm M}_\odot}

\title{Enhancements in velocity-dependent dark matter annihilation in Galactic subhalos}

\author[a, b]{Odelia V. Hartl,}\note{Corresponding author.}
\author[b]{Evan Vienneau,}
\author[a]{Evan Batteas,}
\author[a]{Addy J. Evans,}
\author[b, c]{Nassim Bozorgnia,}
\author[a]{Louis E. Strigari,}

\affiliation[a]{Department of Physics and Astronomy,
Mitchell Institute for Fundamental Physics and Astronomy,
Texas A$\&$M University, College Station, TX 77843, USA}
\affiliation[b]{Department of Physics, University of Alberta,
CCIS 4-181, Edmonton, Alberta T6G 2E1, Canada}
\affiliation[c]{Theoretical Physics Institute, University of Alberta,
CCIS 4-181, Edmonton, Alberta T6G 2E1, Canada}

\emailAdd{odeliah@tamu.edu}

\abstract{We examine velocity-dependent dark matter annihilation in subhalos using a sample of six Milky Way-like galaxies from the Aurgia simulation suite. We quantify the enhancement in the annihilation rate in subhalos when including the contribution from particles in the smooth component of the halo that overlap with the subhalos. The enhancement in the annihilation rate scales with the smooth component of the host halo dark matter density, and is evident for subhalos over the resolvable mass range. Maximal enhancement factors are $\sim 48$ for p-wave models, and $\sim 37,000$ for d-wave models. For p and d-wave annihilation models, $\sim 13$ and $\sim 6$ subhalos, respectively, across all six host halos have emission from dark matter annihilation in their direction that is above the foreground emission from the smooth dark matter component, and would therefore be resolvable as sources. Such subhalos with the most significant enhancement factors tend to be on the lower end of the mass range and located closer to the center of the host galaxy. We provide a prescription to calculate the enhancement for subhalos as a function of distance from the Galactic center, and use this to examine the impact on dark matter limits from a couple of example dwarf spheroidals. We show that, including the enhancement factors, limits from individual dwarf spheroidals are at a cross section scale that may approach those derived from the Galactic center. 
} 

\begin{document}
\maketitle
\flushbottom

\section{Introduction}
\label{sec:intro}

\par Though many sources have been investigated and there still linger some unresolved anomalies, there is no clear evidence for dark matter annihilation from an astrophysical source~\cite{Funk:2013gxa,Conrad:2015bsa,Slatyer:2017sev}. This has led to strong bounds on the standard velocity-independent annihilation cross section, $\sigma_A v_{\rm rel}$, through searches from dwarf spheroidal galaxies (dSphs)~\cite{Fermi-LAT:2011vow,Fermi-LAT:2015att,2024PhRvD.109f3024M}, from the Galactic center~\cite{Abazajian:2020tww,Murgia:2020dzu}, as well as from a variety of other sources. In absence of a definitive signal, a promising direction is to consider unassociated sources in the Fermi-LAT point source catalog as a possible dark matter source~\cite{Fermi-LAT:2012fij,20224FGL3DR}. Indeed, although some unidentified Fermi-LAT sources may be fit by annihilation of dark matter within subhalos, such sources are difficult to definitively model~\cite{Coronado-Blazquez:2019puc,Coronado-Blazquez:2021amj,Gammaldi:2022wwz,Aguirre-Santaella:2023sww}. When considering low-mass subhalos and dwarf galaxies, there are also strong bounds on Sommerfeld models in which the cross section scales as the inverse of the dark matter relative velocity, $1/v_{\rm rel}$~\cite{Kuhlen:2009kx,Boddy:2017vpe}. 

\par Given that velocity-independent s-wave models and Sommerfeld models are becoming strongly constrained, it is important to consider the prospects for detection of more general velocity-dependent dark matter annihilation, which are equally well-motivated but significantly less constrained. Bounds on the annihilation cross section for p and d-wave models, in which $\sigma_A v_{\rm rel}$ scales with the dark matter relative velocity as $v_{\rm rel}^2$ and $v_{\rm rel}^4$, respectively, are much weaker than the corresponding limits from velocity-independent s-wave models. Because of these different velocity scalings, extraction of the signal requires more detailed modeling of the sources as compared to the typical velocity-independent cases~\cite{Smyth:2021bcp,Baxter:2022dpn}. The most stringent limits on the velocity-dependent cross section use the kinematic data from dSphs combined with simplified models for the dark matter velocity distribution~\cite{Boddy:2017vpe,Petac:2018gue,Boddy:2019qak}. These models for the velocity distribution may be compared to those from simulations~\cite{Board:2021bwj,Blanchette:2022hir,Piccirillo:2022qet}, which provide a more accurate representation of the dark matter relative velocity distribution. 

\par Considering dSphs and subhalos within the Milky Way halo, velocity-dependent annihilation models lead to particularly unique phenomenology. For example, it has been recently pointed out that for the particular case of the Sagittarius dSph, there is an enhancement factor to the velocity-dependent annihilation rate which arises because the host subhalo is embedded in the Milky Way halo~\cite{2024JCAP...10..019V}. This effect traces back to the difference in the density and velocity distributions for particles bound to the main Milky Way halo but spatially ovelapping with Sagittarius, and those particles bound to the subhalo hosting Sagittarius. For the sample of Milky Way-like halos and Sagittarius-like galaxies studied, this enhancement factor was found to be as large as a factor of $\sim 30$ for p-wave models relative to the case in which only particles bound to Sagittarius are accounted for, while for d-wave models the enhancement can be over three orders of magnitude. This enhancement increases the sensitivity to p and d-wave annihilation models, deriving from the coarse-grained distributions and is purely astrophysical in nature.  

\par However, there exist some complications for extraction of the enhanced emission component. In particular because of its location on the sky, for Sagittarius the emission from foreground dark matter annihilation within the Milky Way halo was found to be significant, possibly similar to or larger than emission from the subhalo itself. It was not certain that such an effect from the Sagittarius dSph could be disentangled from the foreground. In addition, this foreground emission component has not been accounted for in studies that used dSphs to constrain p and d-wave emission models~\cite{Boddy:2019qak}. 

In this paper, we expand and consider the enhancement factor to p and d-wave models for the entire population of subhalos in a Milky Way-like galaxy. We in particular quantify the enhancement as a function of subhalo properties, such as mass and galactocentric distance, and quantify the mean and scatter of enhancement factors in subhalos. We further quantify the probability of detecting the signal above the foreground dark matter annihilation signal, and identify subhalos that would be detectable above the foreground. We then consider the implications of the enhancement for cross section limits from two example dSphs. 

This paper is organized as follows. In Section~\ref{sims}, we describe the theory and the simulations used in our analysis. In Section~\ref{sec:VBUB_NBUB}, we describe in more detail how unbound dark matter particles affect the annihilation cross section. In Section~\ref{sec:jfactor}, we present the resulting ${\cal J}$-factors for the subhalos from the simulations, and in Section~\ref{sec:gammaray} we discuss the results in the context of gamma-ray bounds from dwarf galaxies. In Section~\ref{sec:discussion}, we end with a discussion and conclusions. 

\section{Simulations and ${\cal J}$-factors}
\label{sims}
In this section we review the simulations, subhalo selection, and theory behind the calculations of the ${\cal J}$-factors. 

\subsection{The Auriga simulations and subhalo selection}

We utilize the Auriga suite of zoom-in simulations  of Milky Way analogue host halos~\cite{Grand:2016mgo, Grand:2024xnm}. These halos were selected from the large dark matter-only EAGLE simulation~\cite{Schaye:2014tpa, 2015MNRAS.450.1937C}, which evolved in a co-moving box size of $\mathrm{(100 \, cMpc)^{3}}$ with cosmological parameters in agreement with Planck-2015~\cite{2016A&A...594A..13P}: $\Omega_{m} =0.307$, $\Omega_{b} =0.048$, $\mathrm{H_{0} = 67.77 \, km \, s^{-1} \, Mpc^{-1}}$. These simulations were run using the moving-mesh code Arepo and implement a  galaxy formation subgrid model, which incorporates star formation, AGN and supernova feedback, black hole formation, metal cooling and background UV/X-ray photoionisation radiation~\cite{Grand:2016mgo}. We use six Milky Way analogues at the high resolution level (Level 3) of the Auriga simulations with dark matter particle mass of $5 \times 10^{4} \, \Msun$, stellar particle mass of $5 \times 10^{3} \, \Msun$ and a Plummer equivalent gravitational softening length, $\epsilon= 184$~pc~\cite{Power:2002sw, Jenkins:2013raa}. 

Subhalos and the bound dark matter particles associated with them are identified using the SUBFIND algorithm~\cite{Springel:2000qu}. In order to sufficiently resolve the density and velocity structure of the subhalos, we utilize only the subhalos that have more than 500 dark matter particles within their virial radius,  $R_{200}$\footnote{The virial radius,  $R_{200}$, is defined as the radius of a sphere centered on the subhalo's center that contains a mean matter density 200 times the critical density of the Universe.}. Of these identified subhalos, we keep only those whose centers of mass and SUBFIND positions are separated by a distance less than 20\% of their $R_{\rm max}$, defined as the radius at which the circular velocity of the subhalo peaks. We choose a 20\% threshold to ensure that any discrepancy between the SUBFIND positions and the center of mass is not significant relative to the sizes of the subhalos. Among the six host Milky Way analogues, we find a total of 902 subhalos that satisfy these criteria. The nearest subhalo to its host galaxy is $\sim 15.7$~kpc from the host center. The mass of the least massive subhalo is $6.12 \times 10^{6}~\Msun$. Note that $\approx 2/3$ of the subhalos in our sample do not have baryons associated with them. 

To approximate the distance of each subhalo from the Sun, we need to specify the Solar position in the simulations. We define a reference frame where the origin is at the  center of the host Milky Way halo and the $z$-axis is in the direction of the stellar disk angular momentum. We then choose the Solar position to be in the stellar disk at  a radial distance of 8 kpc from the center of the host halo and at a random azimuthal angle.

\subsection{${\cal J}$-factors from simulations}
The $\mathcal{J}$-factors for general velocity-dependent annihilation models depend  on both the dark matter density profile and the dark mattter relative velocity distribution of the subhalos. We follow closely the formalism in Refs.~\cite{Board:2021bwj, Blanchette:2022hir} to compute the $\mathcal{J}$-factors for each of the annihilation cross section models, and Ref.~\cite{2024JCAP...10..019V} for the calculation of the contribution from the bound and unbound particles in the subhalos. Regarding the classification as bound and unbound components, for each subhalo we extract the position vector, ${\bf x}$, and the velocity vector, ${\bf v}$, of particles with respect to the center of the subhalo. The bound dark matter particles are those identified by the SUBFIND algorithm as belonging to the given subhalo. The unbound dark matter particles are those belonging to the host Milky Way analogue, but that are within a spherical radius of $R_{\rm max}$ of the subhalo, and are not identified as members of the subhalo via the SUBFIND algorithm. 

The dark matter annihilation cross section averaged over the relative velocity distribution at position $\textbf{x}$ in a subhalo is given by
\begin{equation}
\langle \sigma_A v_{\rm rel} \rangle (\textbf{x}) = \int d^3 \textbf{v}_{\rm rel} P_\textbf{x}(\textbf{v}_{\rm rel})(\sigma_A v_{\rm rel}),
\label{eq:annihilation-cross-section}
\end{equation}
where $P_\textbf{x}(\textbf{v}_{\rm rel})$ is the distribution of
dark matter relative velocities at position ${\bf x}$. In general, $\sigma_A v_{\rm rel}$ can be parametrized as $\sigma_A v_{\rm rel} = (\sigma_A v_{\rm rel})_0 (v_{\rm rel}/c)^n$, and depends on the relative dark matter velocity. Here $(\sigma_A v_{\rm rel})_0$ is the velocity-independent component of the annihilation cross section, and $n$ depends on the specific dark matter annihilation model. We consider the following models: $n=0$ (s-wave annihilation), $n=2$ (p-wave annihilation), $n=4$ (d-wave annihilation), and $n=-1$ (Sommerfeld-enhanced annihilation). From the cross section, the effective $\mathcal{J}$-factor is defined as~\cite{Boddy:2019wfg, Board:2021bwj},
\begin{equation} 
{\mathcal J} (\theta) = \int d \ell \, \frac{\langle \sigma_A v_{\rm rel} \rangle}{(\sigma_A v_{\rm rel})_0}  \left[\rho (r(\ell, \theta))\right]^2 = \int d \ell \int d^3 {\bf v}_{\rm rel} P_{{\bf x}} ({\bf v}_{\rm rel}) ~\left(\frac{{v}_{\rm rel}}{c}\right)^n~ \left[\rho (r(\ell, \theta))\right]^2\, ,
\label{eq:Jfactor}
\end{equation} 
where $\rho(r)$ is the dark matter density profile of the subhalo, $r$ is the radial distance from the subhalo center, $\ell$ is the line of sight from the Sun to a point in the subhalo, and $\theta$ is the opening angle between the line of sight and the distance from the Sun to the subhalo center. The $\mathcal{J}$-factor integrated over solid angle is given by
\begin{equation}
 \widetilde{\mathcal{J}}(\theta)=2\pi \int_0^\theta \mathcal{J}(\theta') \sin \theta' d\theta' . 
 \label{eq:integrated-J}
\end{equation}

We separate the calculation of the $\mathcal{J}$-factors for the smooth halo (which is separate from the unbound particle contribution described above) and subhalos. For the smooth halo component, the local dark matter density around each dark matter particle is estimated using a Voronoi tessellation. Then the dark matter relative velocity distribution is computed at each point on a spherical mesh using the nearest 500 dark matter particles~\cite{Piccirillo:2022qet}. These values are then interpolated to obtain an estimate for the dark matter relative velocity distribution at every point in the smooth halo. The $\mathcal{J}$-factors are then computed by integrating the contribution from all dark matter particles along the line-of-sight to the subhalo. 

For subhalos, the local dark matter density is found using the best fit Einasto density profile within $R_{\rm max}$ and a Voronoi tessellation outside of $R_{\rm max}$. The density profile is computed such that each shell has 100 dark matter particles. The dark matter relative velocity distribution is also calculated in non-uniform consecutive spherical shells from the center of the subhalo, where each shell contains 100 dark matter particles. This allows us to capture the radial dependence of the dark matter relative velocities in the subhalo. The resulting velocity distribution is then used to compute the effective $\mathcal{J}$-factors using Eq.~\eqref{eq:Jfactor}. 

The expected gamma-ray flux from dark matter annihilation is proportional to the ${\mathcal J}$-factor and given by
\begin{equation}
\frac{d \Phi_\gamma}{dE} = \frac{\left(\sigma_A v_{\rm rel}\right)_0}{8 \pi m_{\rm DM}^2}\frac{dN_\gamma}{dE} \,\mathcal{J},
\label{eq:flux}
\end{equation}
where $m_{\rm DM}$ is the dark matter particle mass, and $dN_\gamma/dE$ is the gamma-ray energy spectrum produced per annihilation.
 
\section{Enhancements in density and velocity} 

\label{sec:VBUB_NBUB}

\begin{figure*}
\centering 
    \includegraphics[width = \textwidth]{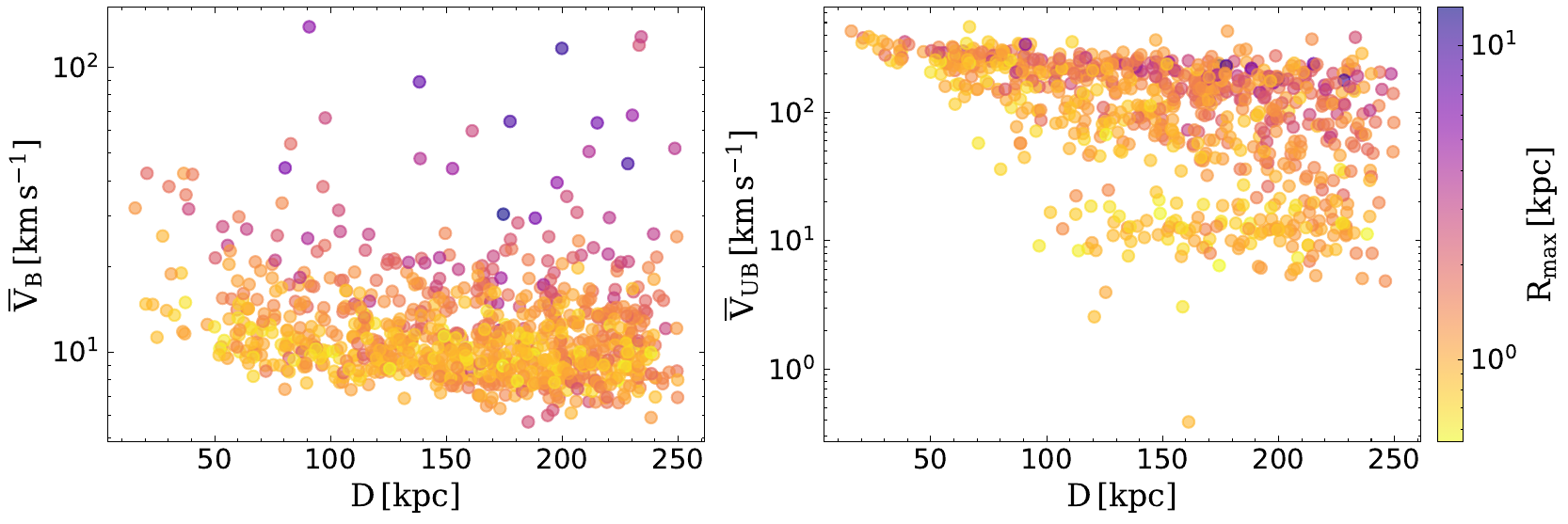}
    \\
    \vspace{2pt}
    \includegraphics[width = .5
\textwidth]{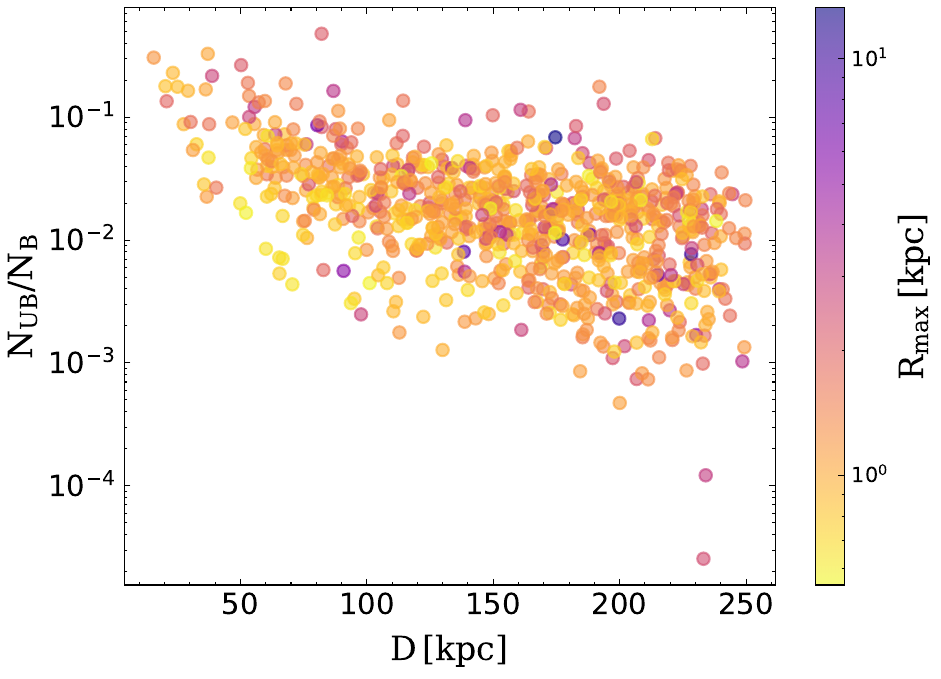}
    
    \caption{The top panels show the relationship between the mean of pairwise velocity distributions for bound (top left) and unbound (top right)  dark matter particles  within the radius $R_{\mathrm{max}}$ of the subhalos and  the distance ${\rm D}$ of the subhalo center to the associated Milky Way analogue host halo. The bottom panel shows the ratio of the total number of unbound dark matter particles within $R_{\mathrm{max}}$ to the number of bound dark matter particles as a function of the distance ${\rm D}$. In all panels, the color bar indicates the size of the subhalos in terms of $R_{\mathrm{max}}$.}
    \label{fig:vel_UB_B}
\end{figure*}

The analysis above shows that the primary factors influencing the ${\cal J}$-factor are the pairwise dark matter velocity distribution and the dark matter density. In Figure~\ref{fig:vel_UB_B}, we explore the relationship between these quantities and the distance of subhalos from the center of their host halo. In the top left panel, we present the mean of the dark matter pair-wise velocity modulus distribution, $\overline V_B$, for the  bound dark matter particles of each subhalo. In general, we do not see a trend between $\overline V_B$  and the galactocentric distance of the subhalo. However, there is a trend indicating that subhalos with larger $R_{\mathrm{max}}$ exhibit higher $\overline V_B$, as seen from the color bar. This is expected since the mean modulus velocity scales with the velocity dispersion of bound dark matter particles in the subhalo, the latter of which scales with subhalo mass. 

The top right panel of Figure~\ref{fig:vel_UB_B} shows the mean of the pair-wise velocity modulus distribution for the unbound dark matter particles, defined as $\overline V_{\rm UB}$. In this case, a clear trend emerges, demonstrating that subhalos closer to the center of their associated host halo have larger $\overline V_{\rm UB}$. This trend simply reflects the dark matter density profile of the host Milky Way halo, in that a larger fraction of unbound particles are associated with subhalos that are closer to the center of the host halo. Additionally, the velocity of particles in the Milky Way increases with decreased distance to the Milky Way center, leading to higher velocities for unbound dark matter particles in subhalos closer to the center. Finally, the bottom panel of Figure~\ref{fig:vel_UB_B} shows the ratio of the number of unbound dark matter particles in the subhalo, $N_{\rm UB}$, to the number of bound dark matter particles, $N_{B}$. An expected relationship is seen for subhalos in closer proximity to the center of the host halo having a larger ratio of unbound dark matter particles to bound particles. As stated above, this is due to the dark matter density profile of the Milky Way, which leads to larger number of unbound particles for subhalos closer to the host center.

From the results in Figure~\ref{fig:vel_UB_B} we can predict the enhancement in annihilation rates for the different velocity dependent cross sections when including the unbound particle component. The enhancement is calculated by integrating the ratio between the $\mathcal{J}$-factor map that is computed with both the bound and unbound dark matter particles and the $\mathcal{J}$-factor map computed with just the bound dark matter particles. The integration is performed over the entire subhalo, defined by the farthest bound particle from the center of the subhalo. In the case of p and d-waves, their dependence on pairwise relative velocities of $v_{\rm rel}^{2}$ and $v_{\rm rel}^{4}$, respectively, suggests these will be enhanced when the unbound particles increase the pair-wise velocity. This may be read directly from the  top right panel of Figure~\ref{fig:vel_UB_B}, which shows that the pair-wise velocities of unbound particles increase with decreased distance of the subhalo to the Milky Way analogue host halo. Therefore, subhalos in closer proximity to the Milky Way analogue would yield a greater enhancement in both the p and d-wave models. In the case of s-wave and Sommerfeld models, the ${\cal J}$-factors are primarily driven by the dark matter density. As indicated above and shown in the bottom panel of Figure~\ref{fig:vel_UB_B}, subhalos situated closer to the Milky Way analogue contain a higher number of unbound particles than  those located at larger radii. Consequently, we can predict that the nearby subhalos will also exhibit a significant boost in s-wave and Sommerfeld models.

\section{${\cal J}$-factor results} 
\label{sec:jfactor}
We now move on in this section to calculate the enhancement factors for velocity-dependent models, and to compare the emission from dark matter annihilation in subhalos to that from the smooth Milky Way halo. 

\subsection{Enhancement when including both bound and unbound particles}

\begin{figure*}
\centering 
    \includegraphics[width = 0.49\textwidth]{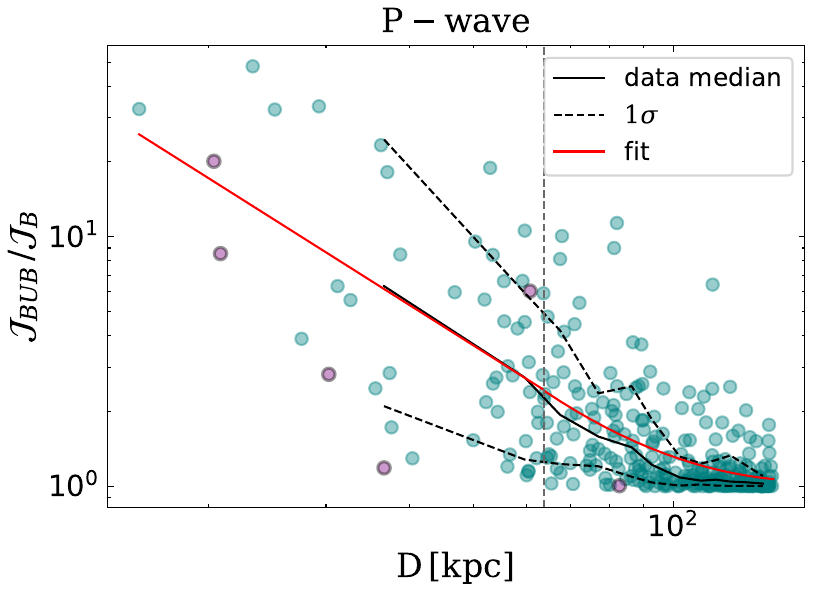}
    \includegraphics[width = 0.49\textwidth]{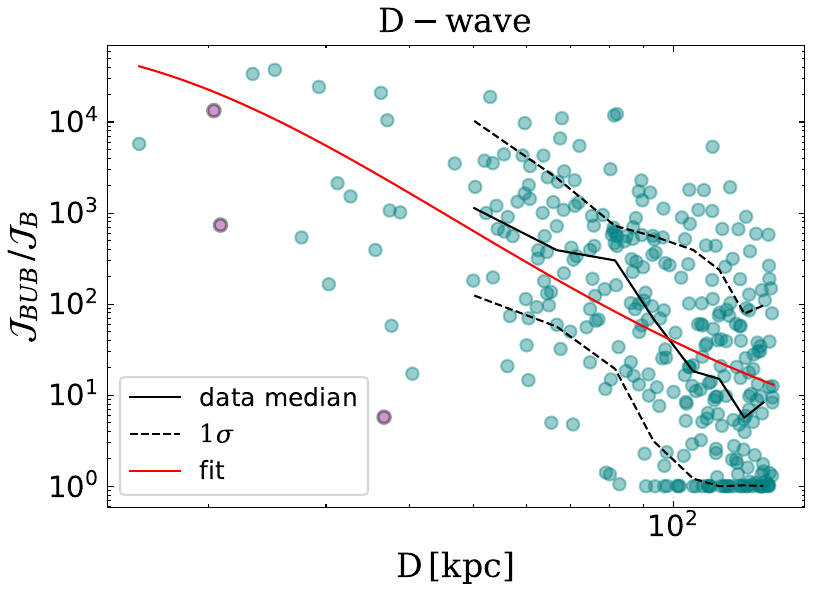}

    \caption{The enhancement in the $\mathcal{J}$-factor  of each subhalo due to the unbound dark matter particles, defined as the ratio of the integrated $\mathcal{J}$-factor for both the bound and unbound dark matter particles, $\mathcal{J}_{\rm BUB}$, and the integrated $\mathcal{J}$-factor including only the bound dark matter particles, $\mathcal{J}_{\rm B}$. This ratio is plotted against the distance, $\mathrm{D}$, from each given subhalo to the center of its associated Milky Way host halo. The solid and dashed black curves show the median enhancement and the $1\sigma$ containment around the median, respectively. The red curves show the fit to the data using the functions and parameters in Eqs.~\eqref{eq:pfit}--\eqref{eq:dfit}. In the left panel, the vertical dashed line shows the break radius used in the fit. The pink points indicate those subhalos that have emission from dark matter annihilation larger than the Milky Way host foreground in their direction and have a non-negligible enhancement, i.e.~$\mathcal{J}_{\rm BUB}/\mathcal{J}_{\rm B}>1$. 
    }
    \label{fig:boost_fit}
\end{figure*}

In Figure~\ref{fig:boost_fit}, we present the enhancement factors as a function of subhalo distance from the galactic center for both p and d-wave models. The enhancement factor is defined as the ratio of the integrated $\mathcal{J}$-factor  for both bound and unbound dark matter particles in the subhalo, $\mathcal{J}_{\rm BUB}$, and the integrated $\mathcal{J}$-factor including only the bound dark matter particles, $\mathcal{J}_{\rm B}$. In agreement with our predictions stated in the previous section, we observe that the enhancement factor increases for subhalos closer to the galactic center.

We bin the sample by the subhalos' separation from the galactic center, such that there are an even number of subhalos, 25 for p-wave and 40 for d-wave, in each bin. To determine the most probable relationship between the enhancement factor and subhalo distance from the host, we calculate the median enhancement in each bin, which is shown as the solid black curve in both panels of Figure~\ref{fig:boost_fit}. These black curves are plotted at the median radius of each bin. The dashed black curves represent the $1\sigma$ containment around the median. 

To better quantify the relationship between the $\mathcal{J}$-factor enhancement and the distance of the subhalos from the Milky Way host, we fit the data presented in Figure~\ref{fig:boost_fit} using a piecewise function for the p-wave and a half-Gaussian function for the d-wave. Here D (in units of  kpc) represents the distance between the subhalo and its associated Milky Way host center, and $\mathcal{J}_{\rm BUB}/\mathcal{J}_{\rm B}$ represents the enhancement of the $\mathcal{J}$-factor for each model due to the inclusion of unbound dark matter particles. For p-wave, the function is specifically 
\begin{equation}
\begin{cases} 
      \log{(\mathrm{\mathcal{J}_{\rm BUB}}/\mathrm{\mathcal{J}_{\rm B}})} = a\log^{2}{\mathrm{D}} + b\log{\mathrm{D}} + c & \log{\mathrm{D}> 1.83} \\
      \log{(\mathrm{\mathcal{J}_{\rm BUB}}/\mathrm{\mathcal{J}_{\rm B}})} = d\log{\mathrm{D}} + e & \log{\mathrm{D}\leq 1.83}. \\      
\end{cases}
\label{eq:pfit}
\end{equation} 
We identify the best-fit parameters by minimizing the $\chi^{2}$ statistic. The corresponding best-fit parameters in each of the cases are 
\begin{equation}
\begin{cases} 
      a = 2.2, \, b = -9.8, \, c = 10.9 \\
      d = 1.7, \, e = 3.4\\       
\end{cases}
\end{equation} 
corresponding to a reduced $\chi^{2}$ of 0.72.
For d-wave, we use a half-Gaussian function,
\begin{equation}
\log{\left(\mathrm{\mathcal{J}_{\rm BUB}}/\mathrm{\mathcal{J}_{\rm B}}\right)}= A\exp{\left[(\log{D}-\mu)^{2}/(2\sigma^{2})\right]}\,,
\label{eq:dfit}
\end{equation} 
with best-fit parameters $A = 4.8, \, \mu = 1, \, \sigma = 0.67$ and a reduced $\chi^{2}$ of 0.67. The best-fit functions are represented by the red curves in Figure \ref{fig:boost_fit}, computed from all points without binning. Consequently, we can use these relationships to predict the enhancement for both the p-wave and d-wave models for a subhalo located at a specified distance.

We can make a rough comparison to the results found in Ref.~\cite{2024JCAP...10..019V}, by taking the average distance of the six Sagittarius analogues identified in Ref.~\cite{2024JCAP...10..019V} to be 42.78 kpc. Input of this distance into our fitting functions yields an enhancement of $\sim 5$ and $\sim 1722$ for the p and d-wave models, respectively. This is in agreement with the results shown in  Figure 6 of Ref.~\cite{2024JCAP...10..019V} for the average enhancement in the integrated $\mathcal{J}$-factor of Sagittarius analogues.

\subsection{Observing subhalos over the smooth Milky Way emission}

Since the smooth component of the halo becomes more significant for p and d-wave models~\cite{Blanchette:2022hir,Piccirillo:2022qet}, it is important to quantity the contribution of dark matter annihilation from the foreground in the emission region of each subhalo. More specifically, even though the emission from dark matter annihilation in subhalos is enhanced in p and d-wave models, is this emission visible above the foreground emission? 

To quantify the foreground emission relative to the subhalo emission, we consider a simple metric comparing the emission from the respective components. To specify the sample of subhalos that have emission above that of the Milky Way foreground, we compute the $\mathcal{J}$-factor as a function of the opening angle from the center of the subhalo. In particular, we integrate over the central region of the subhalo up to an angle of $\sim 0.5$ degrees, which approximately aligns with the characteristic angular resolution of Fermi-LAT~\cite{2012ApJS..203....4A} over its energy range, and identify the subhalos that exhibit emissions above the Milky Way foreground emission.

\begin{figure}[t]
	\centering
	\begin{subfigure}{0.49\linewidth}
		\includegraphics[width=\linewidth]{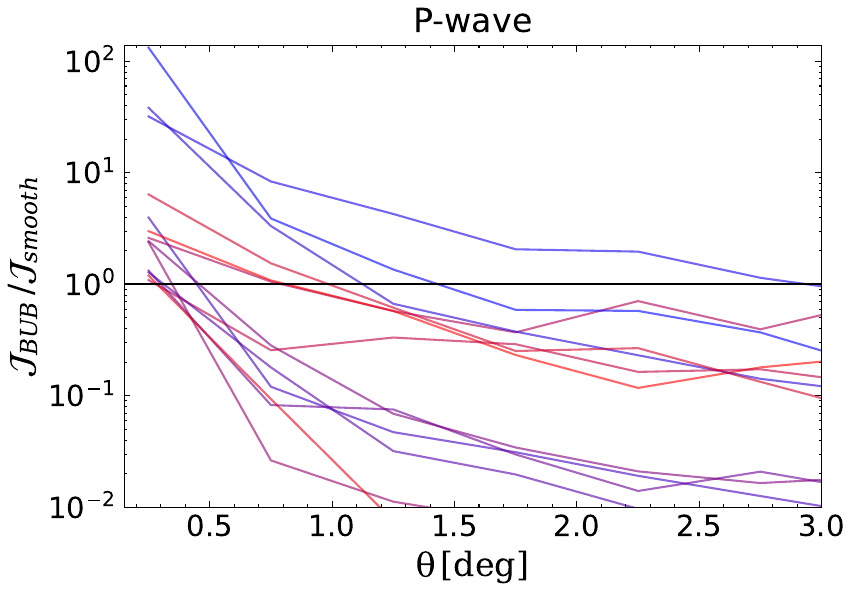}
	
		\label{fig:ubfigA}
	\end{subfigure}
	\begin{subfigure}{0.49\linewidth}
		\includegraphics[width=\linewidth]{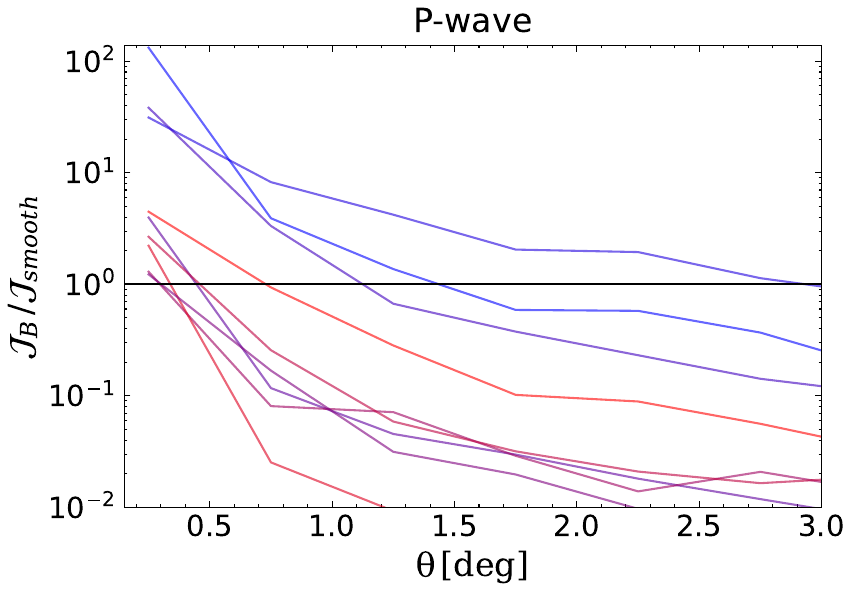}
		
		\label{fig:subfigB}
	\end{subfigure}
	\begin{subfigure}{0.49\linewidth}
	        \includegraphics[width=\linewidth]{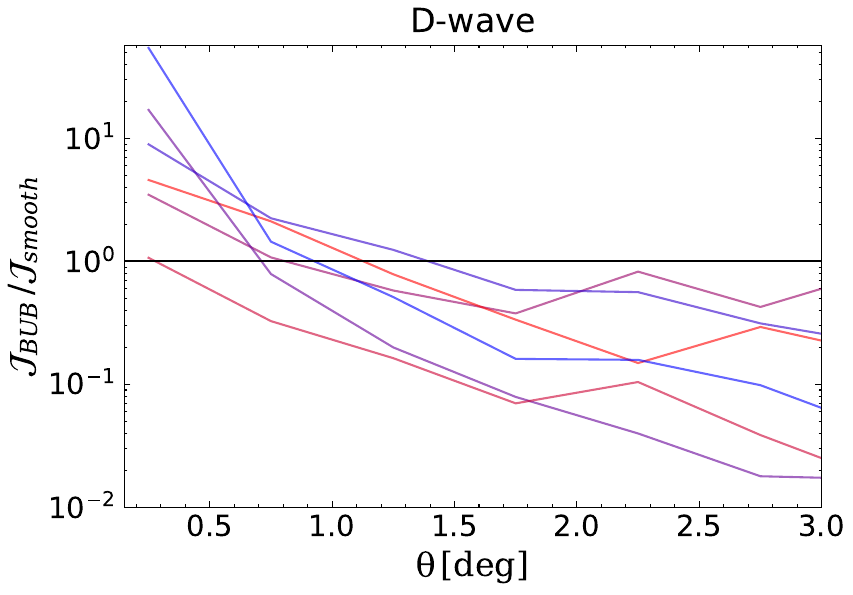}
	        
	        \label{fig:subfigC}
         \end{subfigure}
    \begin{subfigure}{0.49\linewidth}
	        \includegraphics[width=\linewidth]{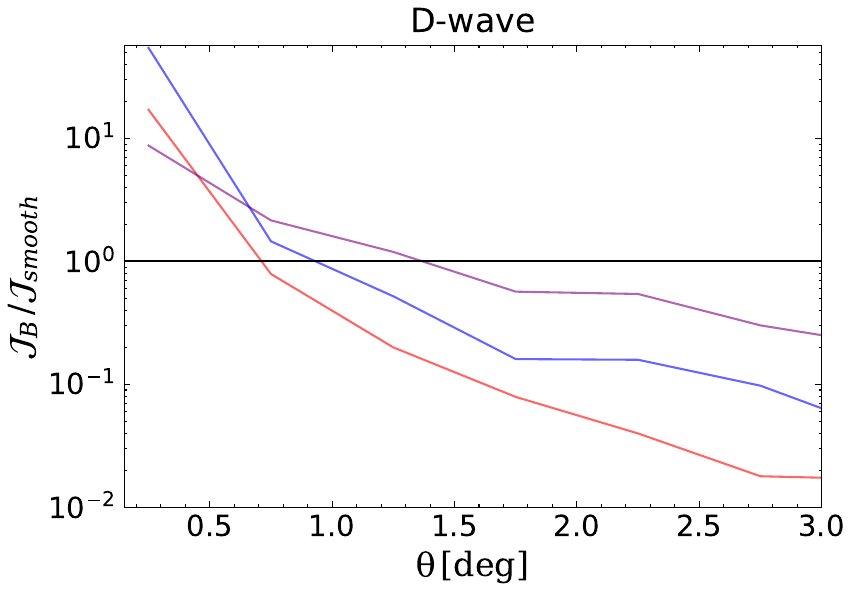}
	       
	        \label{fig:subfigC}
         \end{subfigure}
	\caption{The ratio of the $\mathcal{J}$-factors for selected subhalos and the smooth component of their host Milky Way halo in the direction to the subhalo for the p-wave (top panels) and d-wave (bottom panels) models. The subhalo $\mathcal{J}$-factor includes both bound and unbound dark matter particles in the left panels, while it includes only the bound particles in the right panels. For the case of p-wave, 13 subhalos are detected above the smooth foreground when including both bound and unbound particles (top left), while  9 subhalos are detected above the smooth foreground when including only the bound particles (top right). For d-wave, 6 subhalos are detected above the smooth foreground when both bound and unbound particles are included (bottom left), while this number reduces to 3 when only the bound particles are included. In all panels, the curves are colored by the  dark matter mass of the subhalo, with the color gradient ranging from red to blue representing smaller to larger masses, respectively.
    }
	\label{fig:JBUB}
\end{figure}

We show the ratio of the $\mathcal{J}$-factors for this sample of subhalos and the smooth component of their host Milky Way halo in the subhalo direction  in Figure~\ref{fig:JBUB}. The left panels of the figure show this ratio when both the bound and unbound dark matter particles are included in the calculations of the subhalo $\mathcal{J}$-factor, while in the right panels only the bound dark matter particles are included in the subhalo $\mathcal{J}$-factor. The top and bottom panels show the results for the p and d-wave models, respectively.

We see here that for both p and d-wave, several subhalos have emission in their central regions above the foreground. As listed in Table~\ref{tab:sub_props_all}, the subhalos with enhancements above the foreground span a range of masses and galactocentric distances. For example, we idenfity low-mass subhalos at small galactocentric distances that have a significant enhancement, e.g.~Au16, 140. Further, Figure~\ref{fig:JBUB} shows that the morphology of emission from the subhalos spans a wide range, from subhalos that have extension $\gtrsim 1$ degrees, and those that are more point-like with emission $\lesssim 1$ degrees. The most massive subhalos, on the order of $10^{10} \, \Msun$, have a surprisingly extended emission of $\approx 2$ degrees, which is detected above the foreground. Not surprisingly, the enhancement is negligible for these subhalos due to their high masses and large galactocentric distances, both of which reduce the effect of including the unbound dark matter particles found within the $R_\mathrm{max}$ of the subhalo.

\begin{table*}[t]
	\centering
	\begin{tabular}{|l|c|c|c|c|c|c|} 
            \hline
		Host, Sub & D [kpc] & $M_{\rm DM}$~$[10^8~\Msun]$ & $M_{\star}$~$[10^6~\Msun]$ &$R_\mathrm{max}$~[kpc] & P boost & D boost\\
				\hline
    Au21, 10 & 36.7 & 3.23&279.40 & 1.16 & 1.2 &5.8\\
    Au16, 9 & 20.8 & 4.91&85.68 & 1.93 & 8.6 &738.4 \\
    Au16, 140 & 20.4 & .33&0.54 & 0.84 & 20.0 &13334.1 \\
    \hline
     Au23, 3& 152.6& 17.98 &96.03& 4.52& 1.0& 1.2\\
    Au16, 4 &82.8 &21.78& 350.08 &1.92 &1.0 &1.1 \\
    Au24, 221 &  60.8 &0.51 &0.0& 1.21 &6.1& 3285.0\\
  Au23, 7& 30.3& 4.78& 111.33& 1.72& 2.8& 165.2\\
    \hline
	\end{tabular}
	\caption{Properties of the select subhalos that have emission greater than the foreground in their direction, and have a non-negligible enhancement factor, i.e.~$>1$. The columns from left to right show the host and subhalo IDs, the subhalo distance ${\rm D}$ to the the host halo, the dark matter mass of the subhalo $M_{\rm DM}$, the stellar mass of the subhalo $M_\star$, the radius $R_{\rm max}$ of the subhalo, and the $\mathcal{J}$-factor enhancements for the p-wave and d-wave models. Subhalos above the horizontal line have p-wave emission that makes them detectable as extended sources.}
 \label{tab:sub_props_all}
\end{table*}

\section{Gamma-ray limits} 
\label{sec:gammaray}
We can use our prediction for the velocity-dependent $\mathcal{J}$-factor enhancement and apply it to known dSphs  in the Milky Way. We focus on two ultra-faint dwarf galaxies that are among the nearest to the Galactic center, Ursa Major II and Segue 1. For Ursa Major II, the integrated $\log_{10} [ \mathcal{J} /({\rm GeV}^2~{\rm cm}^{-5}) ]$ within 1 degree is 11.51 and 3.83 for p and d-wave models, respectively, while for Segue 1, the integrated $\log_{10} [ \mathcal{J} /({\rm GeV}^2~{\rm cm}^{-5}) ]$ within 1 degree is 10.21 and 1.57 for p and d-wave models, respectively~\cite{Boddy:2019qak}. Note that these were derived under simplified assumptions of a Maxwellian velocity distribution for the dark matter in the dwarf galaxies, but they are appropriate examples for the purposes of illustrating our point of how the limits are affected. 

The heliocentric distances to Ursa Major II and Segue 1 are 32 and 23 kpc, which for their position in the sky corresponds to a galactocentric distance of $\sim 40$ and $\sim 28$ kpc, respectively. From our fitting formula above, the p-wave enhancement factor for Ursa Major~II is $\sim 5$, while for Segue 1 it is $\sim 10$. The d-wave enhancements for these galaxies are $\sim 1719$ and $\sim 7664$, respectively. We now examine how these enhancement factors affect the bounds on the dark matter annihilation cross section derived from Fermi-LAT. 

To derive updated gamma-ray limits for each galaxy, we utilize $\texttt{FermiPy}$ with Fermitools 2.2.0~\cite{2017ICRC...35..824W}. We use 16.57 years of data, corresponding to mission elapsed times between 239557417 s and 762177465 s. We perform the standard data cuts for analysis of these systems~\cite{McDaniel:2023bju}. We select $\texttt{FRONT}$ and $\texttt{BACK}$ converting (evclass == 128 and evtype  == 3  ) events with energies between 500 MeV and 100 GeV. We apply the suggested $\texttt{(DATA\_QUAL>0) \&\& (LAT\_CONFIG==1) }$ filter to ensure quality data and a zenith cut of $z_{\text{max}} = 90^{\circ}$ to filter out gamma-ray contamination from the Earth's limb.

We consider a $10^{\circ}  \times 10^{\circ}$ ROI centered on each of Ursa Major II and Segue 1. We use the $\texttt{MINUIT}$ optimizer within $\texttt{gtlike}$ with a $0.01^{\circ}$ angular pixelation for our likelihood maximization. For the interstellar emission model we use the recommended $\texttt{gll\_iem\_v07.fits}$, and for the isotropic emission we use $\texttt{iso\_P8R3\_SOURCE\_V3\_v1.txt}$. For simplicity, here we only assume a dark matter annihilation spectrum defined by the FermiPy $\texttt{DmFitFunction}$~\cite{Jeltema:2008hf}, assuming annihilation via the $b\bar{b}$ channel.

\begin{figure*}
\centering 
    \includegraphics[width = \textwidth]{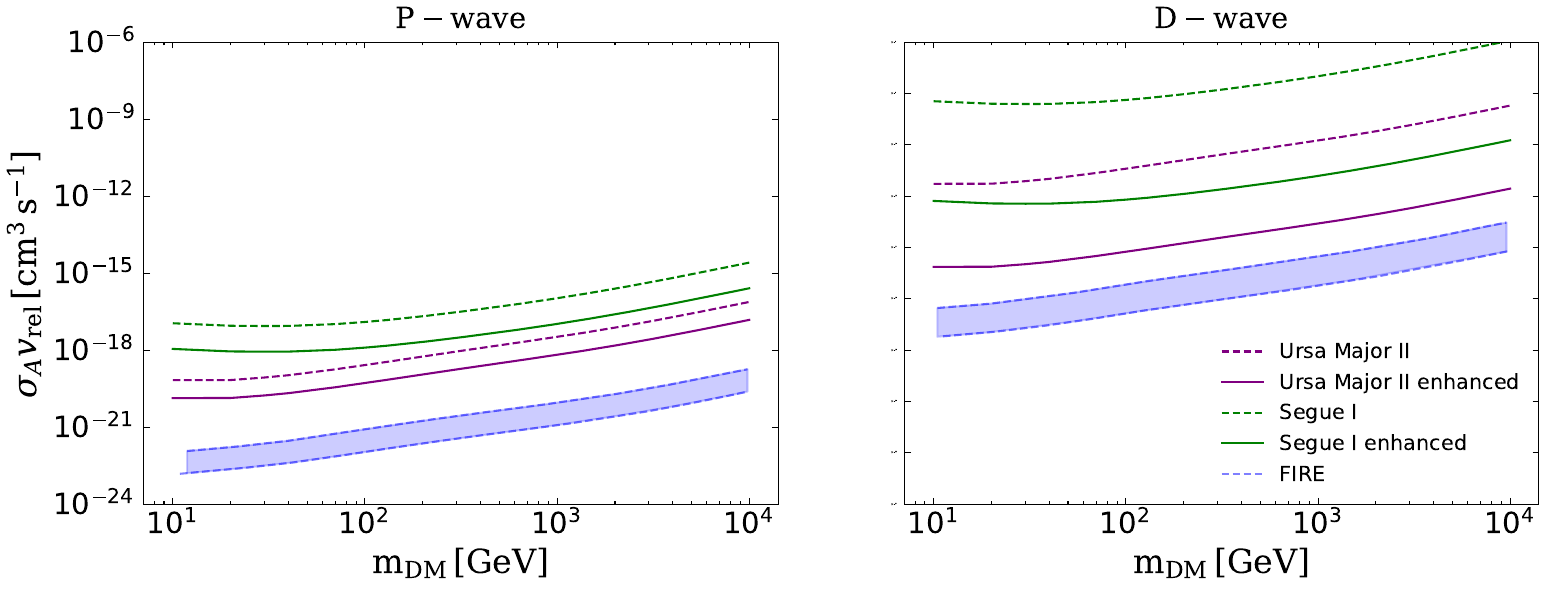}
    \caption{Fermi-LAT gamma-ray upper limits on the dark matter annihilation cross section as a function of the dark matter particle mass from Ursa Major II and Segue 1, for p-wave (left panel) and d-wave (right panel) models. Shown are the scaled upper limits when accounting for the enhancement factors discussed in this paper. The blue band is the range of limits derived from the FIRE simulations using measurements from the Galactic center~\cite{McKeown:2021sob}. 
    }
    \label{fig:upperlimits}
\end{figure*}

Similar to previous studies, we find no detection of significant emission within the regions around the respective dSphs. The resulting 90\% CL upper limits on the dark matter annihilation cross section derived from the null detections are shown in Figure~\ref{fig:upperlimits}. Note that the limits we derive are comparable to the combined limits from previous analyses, though we note that in our case we do not account for uncertainties in the $\mathcal{J}$-factor. We do this for simplicity in interpreting the impact of the velocity-dependent enhancement factors. Figure~\ref{fig:upperlimits} shows how the upper limits scale if we were to implement the enhancement factor predicted from our fit to the simulation data. Here we see the important impact of including the enhancement factor, and how it strengthens the limit on the cross section for all dark matter mass scales. We see that the limits begin to approach the sensitivity derived from the Galactic center using the FIRE simulations~\cite{McKeown:2021sob}. 

\section{Discussion and conclusions}
\label{sec:discussion}
In this paper we have quantified in detail the contributions to the emission from velocity-dependent dark matter annihilation from subhalos in the Milky Way. In particular, we emphasize the importance of the contribution from the dynamical component bound to the Galactic halo, which has a much larger velocity dispersion than the component bound to the subhalo, even though the components overlap spatially. This analysis extends on recent work which quantified this effect in particular for Sagittarius analogues in the Auriga simulations~\cite{2024JCAP...10..019V}. This component has traditionally been neglected in the determination of dark matter annihilation cross section limits, as it not possible to quantify it without sufficiently high resolution simulations.  

We have quantified the behavior of the contribution from the dynamical component bound to the Galactic halo across the entire subhalo population. We identify a relationship between the enhancement factor and the galactocentric distance of a subhalo, which derives from the fall-off of the smooth halo density from the center of the Milky Way. We quantify the median enhancement factor for subhalos at a  galactocentric distance, and provide formulae which may be used for known dSphs. We further identify and characterize a significant scatter in the enhancement factor for subhalos closer to the Galactic center. While the scaling of the enhancement factor with galactocentric distance is straightforward, it is not yet clear what drives the origin of the scatter from subhalo to subhalo. For example, the scatter could be intrinsic, and be a reflection of a physical effect such as dynamical friction which destroys subhalos near the Galactic center. Another possibility is that the scatter simply is a reflection of the noise due to the low sample of subhalos near the Galactic center. This latter component of the scatter, in particular, would be better understood with a larger sample of subhalos from simulated Milky Way-like host systems. 

We have further examined the emission from dark matter annihilation in the subhalos in comparison to the emission from the smooth component of the host Milky Way-like halo in the directions of each of the subhalos. This is especially important for p and d-wave models, for which the emission from the smooth component of the halo tends to dominate that of the subhalo. We have shown, in these models, several subhalos have emission larger than the smooth foreground component, especially when the unbound dark matter particles are included in the calculations. This implies that for all models, appropriate subhalos may be identified to obtain velocity-dependent cross section limits, and the enhancement due to the smooth halo component will increase the predicted emission from the subhalos. 

We find that subhalos that are observable above the foreground and with enhanced emission have a range of spatial extensions, with some extending $\gtrsim 1$ degree, while others exhibit more point-source like emission. The most massive subhalos tend to have the smallest enhancement, primarily because these subhalos are located far away from the Galactic center. The subhalos with the largest enhancements tend to be on the smaller end of the mass spectrum and closest to the Galactic center. These latter subhalos have a range of emission profiles, some being extended and some being more point-like in nature. 

We have performed a Fermi-LAT gamma-ray analysis on two candidate dSphs that have the largest predicted enhancement factors, Ursa Major II and Segue 1. Inclusion of the velocity-dependent enhancement factor strengthens the upper bound on the dark matter annihilation cross section. In addition to the results from individual dwarf galaxies that we have derived, implementation of the enhancement factors that we derive will certainly improve bounds from the entire combined population of dSphs. While beyond the scope of the this paper, such an analysis, when combined with updated data sets from observations, is an important topic for future study, as their sensitivities may rival those from the Galactic center.

\acknowledgments
OH, AE, and LS acknowledge support from DOE Grant de-sc0010813 from the Texas A\&M University System National Laboratories Office and Los Alamos National Laboratory. EV and NB acknowledge the support of the Natural Sciences and Engineering Research Council of Canada (NSERC), funding reference number RGPIN-2020-07138 and the NSERC Discovery Launch Supplement, DGECR-2020-00231. NB acknowledges the support of the Canada Research Chairs Program. We have used simulations from the Auriga Project public data release~\citep{Grand:2024} available at https://wwwmpa.mpa-garching.mpg.de/auriga/data. This work was completed using the \textsc{Python} programming language as well as the following software packages: \textsc{astropy} ~\citep{astropy:2018}, \textsc{pandas}~\citep{reback2020pandas}, \textsc{numpy}~\citep{         harris2020array}, \textsc{scipy}~\citep{2020SciPy-NMeth}, \textsc{matplotlib}~\citep{Hunter:2007}, and \textsc{spyder}~\citep{raybaut2009spyder}.
\bibliographystyle{JHEP}
\bibliography{biblio}

\providecommand{\href}[2]{#2}\begingroup\raggedright\begin{thebibliography}{10}

\bibitem{Funk:2013gxa}
S.~Funk, \emph{{Indirect Detection of Dark Matter with gamma rays}}, \href{https://doi.org/10.1073/pnas.1308728111}{\emph{Proc. Nat. Acad. Sci.} {\bfseries 112} (2015) 2264} [\href{https://arxiv.org/abs/1310.2695}{{\ttfamily 1310.2695}}].

\bibitem{Conrad:2015bsa}
J.~Conrad, J.~Cohen-Tanugi and L.E.~Strigari, \emph{{WIMP searches with gamma rays in the Fermi era: challenges, methods and results}}, \href{https://doi.org/10.1134/S1063776115130099}{\emph{J. Exp. Theor. Phys.} {\bfseries 121} (2015) 1104} [\href{https://arxiv.org/abs/1503.06348}{{\ttfamily 1503.06348}}].

\bibitem{Slatyer:2017sev}
T.R.~Slatyer, \emph{{Indirect detection of dark matter.}},  in \emph{{Theoretical Advanced Study Institute in Elementary Particle Physics}: {Anticipating the Next Discoveries in Particle Physics}}, pp.~297--353, 2018, \href{https://doi.org/10.1142/9789813233348_0005}{DOI} [\href{https://arxiv.org/abs/1710.05137}{{\ttfamily 1710.05137}}].

\bibitem{Fermi-LAT:2011vow}
{\scshape Fermi-LAT} collaboration, \emph{{Constraining Dark Matter Models from a Combined Analysis of Milky Way Satellites with the Fermi Large Area Telescope}}, \href{https://doi.org/10.1103/PhysRevLett.107.241302}{\emph{Phys. Rev. Lett.} {\bfseries 107} (2011) 241302} [\href{https://arxiv.org/abs/1108.3546}{{\ttfamily 1108.3546}}].

\bibitem{Fermi-LAT:2015att}
{\scshape Fermi-LAT} collaboration, \emph{{Searching for Dark Matter Annihilation from Milky Way Dwarf Spheroidal Galaxies with Six Years of Fermi Large Area Telescope Data}}, \href{https://doi.org/10.1103/PhysRevLett.115.231301}{\emph{Phys. Rev. Lett.} {\bfseries 115} (2015) 231301} [\href{https://arxiv.org/abs/1503.02641}{{\ttfamily 1503.02641}}].

\bibitem{2024PhRvD.109f3024M}
A.~{McDaniel}, M.~{Ajello}, C.M.~{Karwin}, M.~{Di Mauro}, A.~{Drlica-Wagner} and M.A.~{S{\'a}nchez-Conde}, \emph{{Legacy analysis of dark matter annihilation from the Milky Way dwarf spheroidal galaxies with 14 years of Fermi -LAT data}}, \href{https://doi.org/10.1103/PhysRevD.109.063024}{\emph{\prd} {\bfseries 109} (2024) 063024} [\href{https://arxiv.org/abs/2311.04982}{{\ttfamily 2311.04982}}].

\bibitem{Abazajian:2020tww}
K.N.~Abazajian, S.~Horiuchi, M.~Kaplinghat, R.E.~Keeley and O.~Macias, \emph{{Strong constraints on thermal relic dark matter from Fermi-LAT observations of the Galactic Center}}, \href{https://doi.org/10.1103/PhysRevD.102.043012}{\emph{Phys. Rev. D} {\bfseries 102} (2020) 043012} [\href{https://arxiv.org/abs/2003.10416}{{\ttfamily 2003.10416}}].

\bibitem{Murgia:2020dzu}
S.~Murgia, \emph{{The Fermi\textendash{}LAT Galactic Center Excess: Evidence of Annihilating Dark Matter?}}, \href{https://doi.org/10.1146/annurev-nucl-101916-123029}{\emph{Ann. Rev. Nucl. Part. Sci.} {\bfseries 70} (2020) 455}.

\bibitem{Fermi-LAT:2012fij}
{\scshape Fermi-LAT} collaboration, \emph{{Search for Dark Matter Satellites using the FERMI-LAT}}, \href{https://doi.org/10.1088/0004-637X/747/2/121}{\emph{Astrophys. J.} {\bfseries 747} (2012) 121} [\href{https://arxiv.org/abs/1201.2691}{{\ttfamily 1201.2691}}].

\bibitem{20224FGL3DR}
{Fermi-LAT collaboration}, {:}, S.~{Abdollahi}, F.~{Acero}, L.~{Baldini}, J.~{Ballet} et~al., \emph{{Incremental Fermi Large Area Telescope Fourth Source Catalog}}, {\emph{arXiv e-prints} (2022) arXiv:2201.11184} [\href{https://arxiv.org/abs/2201.11184}{{\ttfamily 2201.11184}}].

\bibitem{Coronado-Blazquez:2019puc}
J.~Coronado-Blazquez, M.A.~Sanchez-Conde, A.~Dominguez, A.~Aguirre-Santaella, M.~Di~Mauro, N.~Mirabal et~al., \emph{{Unidentified Gamma-ray Sources as Targets for Indirect Dark Matter Detection with the Fermi-Large Area Telescope}}, \href{https://doi.org/10.1088/1475-7516/2019/07/020}{\emph{JCAP} {\bfseries 07} (2019) 020} [\href{https://arxiv.org/abs/1906.11896}{{\ttfamily 1906.11896}}].

\bibitem{Coronado-Blazquez:2021amj}
{\scshape Fermi-LAT} collaboration, \emph{{Spatial extension of dark subhalos as seen by Fermi-LAT and implications for WIMP constraints}},  \href{https://arxiv.org/abs/2204.00267}{{\ttfamily 2204.00267}}.

\bibitem{Gammaldi:2022wwz}
V.~Gammaldi, B.~Zald\'\i{}var, M.A.~S\'anchez-Conde and J.~Coronado-Bl\'azquez, \emph{{A search for dark matter among Fermi-LAT unidentified sources with systematic features in machine learning}}, \href{https://doi.org/10.1093/mnras/stad066}{\emph{Mon. Not. Roy. Astron. Soc.} {\bfseries 520} (2023) 1348} [\href{https://arxiv.org/abs/2207.09307}{{\ttfamily 2207.09307}}].

\bibitem{Aguirre-Santaella:2023sww}
A.~Aguirre-Santaella and M.A.~S\'anchez-Conde, \emph{{The viability of low-mass subhaloes as targets for gamma-ray dark matter searches}}, \href{https://doi.org/10.1093/mnras/stae940}{\emph{Mon. Not. Roy. Astron. Soc.} {\bfseries 530} (2024) 2496} [\href{https://arxiv.org/abs/2309.02330}{{\ttfamily 2309.02330}}].

\bibitem{Kuhlen:2009kx}
M.~Kuhlen, P.~Madau and J.~Silk, \emph{{Exploring Dark Matter with Milky Way substructure}}, \href{https://doi.org/10.1126/science.1174881}{\emph{Science} {\bfseries 325} (2009) 970} [\href{https://arxiv.org/abs/0907.0005}{{\ttfamily 0907.0005}}].

\bibitem{Boddy:2017vpe}
K.K.~Boddy, J.~Kumar, L.E.~Strigari and M.-Y.~Wang, \emph{{Sommerfeld-Enhanced $J$-Factors For Dwarf Spheroidal Galaxies}}, \href{https://doi.org/10.1103/PhysRevD.95.123008}{\emph{Phys. Rev. D} {\bfseries 95} (2017) 123008} [\href{https://arxiv.org/abs/1702.00408}{{\ttfamily 1702.00408}}].

\bibitem{Smyth:2021bcp}
N.~Smyth, G.~Huckabee and S.~Profumo, \emph{{Optimal observing strategies for velocity-suppressed dark matter annihilation}}, \href{https://doi.org/10.1103/PhysRevD.104.123003}{\emph{Phys. Rev. D} {\bfseries 104} (2021) 123003} [\href{https://arxiv.org/abs/2105.03438}{{\ttfamily 2105.03438}}].

\bibitem{Baxter:2022dpn}
E.J.~Baxter, J.~Kumar, A.D.~Paul and J.~Runburg, \emph{{Searching for velocity-dependent dark matter annihilation signals from extragalactic halos}}, \href{https://doi.org/10.1088/1475-7516/2022/09/026}{\emph{JCAP} {\bfseries 09} (2022) 026} [\href{https://arxiv.org/abs/2205.02386}{{\ttfamily 2205.02386}}].

\bibitem{Petac:2018gue}
M.~Petac, P.~Ullio and M.~Valli, \emph{{On velocity-dependent dark matter annihilations in dwarf satellites}}, \href{https://doi.org/10.1088/1475-7516/2018/12/039}{\emph{JCAP} {\bfseries 12} (2018) 039} [\href{https://arxiv.org/abs/1804.05052}{{\ttfamily 1804.05052}}].

\bibitem{Boddy:2019qak}
K.K.~Boddy, J.~Kumar, A.B.~Pace, J.~Runburg and L.E.~Strigari, \emph{{Effective $J$-factors for Milky Way dwarf spheroidal galaxies with velocity-dependent annihilation}}, \href{https://doi.org/10.1103/PhysRevD.102.023029}{\emph{Phys. Rev. D} {\bfseries 102} (2020) 023029} [\href{https://arxiv.org/abs/1909.13197}{{\ttfamily 1909.13197}}].

\bibitem{Board:2021bwj}
E.~Board, N.~Bozorgnia, L.E.~Strigari, R.J.J.~Grand, A.~Fattahi, C.S.~Frenk et~al., \emph{{Velocity-dependent J-factors for annihilation radiation from cosmological simulations}}, \href{https://doi.org/10.1088/1475-7516/2021/04/070}{\emph{JCAP} {\bfseries 04} (2021) 070} [\href{https://arxiv.org/abs/2101.06284}{{\ttfamily 2101.06284}}].

\bibitem{Blanchette:2022hir}
K.~Blanchette, E.~Piccirillo, N.~Bozorgnia, L.E.~Strigari, A.~Fattahi, C.S.~Frenk et~al., \emph{{Velocity-dependent J-factors for Milky Way dwarf spheroidal analogues in cosmological simulations}}, \href{https://doi.org/10.1088/1475-7516/2023/03/021}{\emph{JCAP} {\bfseries 03} (2023) 021} [\href{https://arxiv.org/abs/2207.00069}{{\ttfamily 2207.00069}}].

\bibitem{Piccirillo:2022qet}
E.~Piccirillo, K.~Blanchette, N.~Bozorgnia, L.E.~Strigari, C.S.~Frenk, R.J.J.~Grand et~al., \emph{{Velocity-dependent annihilation radiation from dark matter subhalos in cosmological simulations}}, \href{https://doi.org/10.1088/1475-7516/2022/08/058}{\emph{JCAP} {\bfseries 08} (2022) 058} [\href{https://arxiv.org/abs/2203.08853}{{\ttfamily 2203.08853}}].

\bibitem{2024JCAP...10..019V}
E.~{Vienneau}, A.J.~{Evans}, O.V.~{Hartl}, N.~{Bozorgnia}, L.E.~{Strigari}, A.H.~{Riley} et~al., \emph{{Significant impact of Galactic dark matter particles on annihilation signals from Sagittarius analogues}}, \href{https://doi.org/10.1088/1475-7516/2024/10/019}{\emph{\jcap} {\bfseries 2024} (2024) 019} [\href{https://arxiv.org/abs/2403.15544}{{\ttfamily 2403.15544}}].

\bibitem{Grand:2016mgo}
R.J.J.~Grand, F.A.~Gómez, F.~Marinacci, R.~Pakmor, V.~Springel, D.J.R.~Campbell et~al., \emph{{The Auriga Project: the properties and formation mechanisms of disc galaxies across cosmic time}}, \href{https://doi.org/10.1093/mnras/stx071}{\emph{Mon. Not. Roy. Astron. Soc.} {\bfseries 467} (2017) 179} [\href{https://arxiv.org/abs/1610.01159}{{\ttfamily 1610.01159}}].

\bibitem{Grand:2024xnm}
R.J.J.~Grand, F.~Fragkoudi, F.A.~G\'omez, A.~Jenkins, F.~Marinacci, R.~Pakmor et~al., \emph{{Overview and public data release of the augmented Auriga Project: cosmological simulations of dwarf and Milky Way-mass galaxies}}, \href{https://doi.org/10.1093/mnras/stae1598}{\emph{Mon. Not. Roy. Astron. Soc.} {\bfseries 532} (2024) 1814} [\href{https://arxiv.org/abs/2401.08750}{{\ttfamily 2401.08750}}].

\bibitem{Schaye:2014tpa}
J.~Schaye et~al., \emph{{The EAGLE project: Simulating the evolution and assembly of galaxies and their environments}}, \href{https://doi.org/10.1093/mnras/stu2058}{\emph{Mon. Not. Roy. Astron. Soc.} {\bfseries 446} (2015) 521} [\href{https://arxiv.org/abs/1407.7040}{{\ttfamily 1407.7040}}].

\bibitem{2015MNRAS.450.1937C}
R.A.~{Crain}, J.~{Schaye}, R.G.~{Bower}, M.~{Furlong}, M.~{Schaller}, T.~{Theuns} et~al., \emph{{The EAGLE simulations of galaxy formation: calibration of subgrid physics and model variations}}, \href{https://doi.org/10.1093/mnras/stv725}{\emph{\mnras} {\bfseries 450} (2015) 1937} [\href{https://arxiv.org/abs/1501.01311}{{\ttfamily 1501.01311}}].

\bibitem{2016A&A...594A..13P}
{Planck Collaboration}, P.A.R.~{Ade}, N.~{Aghanim}, M.~{Arnaud}, M.~{Ashdown}, J.~{Aumont} et~al., \emph{{Planck 2015 results. XIII. Cosmological parameters}}, \href{https://doi.org/10.1051/0004-6361/201525830}{\emph{\aap} {\bfseries 594} (2016) A13} [\href{https://arxiv.org/abs/1502.01589}{{\ttfamily 1502.01589}}].

\bibitem{Power:2002sw}
C.~Power, J.F.~Navarro, A.~Jenkins, C.S.~Frenk, S.D.M.~White, V.~Springel et~al., \emph{{The Inner structure of Lambda CDM halos. 1. A Numerical convergence study}}, \href{https://doi.org/10.1046/j.1365-8711.2003.05925.x}{\emph{Mon. Not. Roy. Astron. Soc.} {\bfseries 338} (2003) 14} [\href{https://arxiv.org/abs/astro-ph/0201544}{{\ttfamily astro-ph/0201544}}].

\bibitem{Jenkins:2013raa}
A.~Jenkins, \emph{{A new way of setting the phases for cosmological multi-scale Gaussian initial conditions}}, \href{https://doi.org/10.1093/mnras/stt1154}{\emph{Mon. Not. Roy. Astron. Soc.} {\bfseries 434} (2013) 2094} [\href{https://arxiv.org/abs/1306.5968}{{\ttfamily 1306.5968}}].

\bibitem{Springel:2000qu}
V.~Springel, S.D.M.~White, G.~Tormen and G.~Kauffmann, \emph{{Populating a cluster of galaxies. 1. Results at z = 0}}, \href{https://doi.org/10.1046/j.1365-8711.2001.04912.x}{\emph{Mon. Not. Roy. Astron. Soc.} {\bfseries 328} (2001) 726} [\href{https://arxiv.org/abs/astro-ph/0012055}{{\ttfamily astro-ph/0012055}}].

\bibitem{Boddy:2019wfg}
K.K.~Boddy, J.~Kumar, J.~Runburg and L.E.~Strigari, \emph{{Angular distribution of gamma-ray emission from velocity-dependent dark matter annihilation in subhalos}}, \href{https://doi.org/10.1103/PhysRevD.100.063019}{\emph{Phys. Rev. D} {\bfseries 100} (2019) 063019} [\href{https://arxiv.org/abs/1905.03431}{{\ttfamily 1905.03431}}].

\bibitem{2012ApJS..203....4A}
M.~{Ackermann}, M.~{Ajello}, A.~{Albert}, A.~{Allafort}, W.B.~{Atwood}, M.~{Axelsson} et~al., \emph{{The Fermi Large Area Telescope on Orbit: Event Classification, Instrument Response Functions, and Calibration}}, \href{https://doi.org/10.1088/0067-0049/203/1/4}{\emph{\apjs} {\bfseries 203} (2012) 4} [\href{https://arxiv.org/abs/1206.1896}{{\ttfamily 1206.1896}}].

\bibitem{2017ICRC...35..824W}
M.~{Wood}, R.~{Caputo}, E.~{Charles}, M.~{Di Mauro}, J.~{Magill}, J.S.~{Perkins} et~al., \emph{{Fermipy: An open-source Python package for analysis of Fermi-LAT Data}},  in \emph{35th International Cosmic Ray Conference (ICRC2017)}, vol.~301 of \emph{International Cosmic Ray Conference}, p.~824, July, 2017, \href{https://doi.org/10.22323/1.301.0824}{DOI} [\href{https://arxiv.org/abs/1707.09551}{{\ttfamily 1707.09551}}].

\bibitem{McDaniel:2023bju}
A.~McDaniel, M.~Ajello, C.M.~Karwin, M.~Di~Mauro, A.~Drlica-Wagner and M.~S\`anchez-Conde, \emph{{Legacy Analysis of Dark Matter Annihilation from the Milky Way Dwarf Spheroidal Galaxies with 14 Years of Fermi-LAT Data}},  \href{https://arxiv.org/abs/2311.04982}{{\ttfamily 2311.04982}}.

\bibitem{Jeltema:2008hf}
T.E.~Jeltema and S.~Profumo, \emph{{Fitting the Gamma-Ray Spectrum from Dark Matter with DMFIT: GLAST and the Galactic Center Region}}, \href{https://doi.org/10.1088/1475-7516/2008/11/003}{\emph{JCAP} {\bfseries 11} (2008) 003} [\href{https://arxiv.org/abs/0808.2641}{{\ttfamily 0808.2641}}].

\bibitem{McKeown:2021sob}
D.~McKeown, J.S.~Bullock, F.J.~Mercado, Z.~Hafen, M.~Boylan-Kolchin, A.~Wetzel et~al., \emph{{Amplified J-factors in the Galactic Centre for velocity-dependent dark matter annihilation in FIRE simulations}}, \href{https://doi.org/10.1093/mnras/stac966}{\emph{Mon. Not. Roy. Astron. Soc.} {\bfseries 513} (2022) 55} [\href{https://arxiv.org/abs/2111.03076}{{\ttfamily 2111.03076}}].

\bibitem{Grand:2024}
R.J.J.~{Grand}, F.~{Fragkoudi}, F.A.~{G{\'o}mez}, A.~{Jenkins}, F.~{Marinacci}, R.~{Pakmor} et~al., \emph{{Overview and public data release of the Auriga Project: cosmological simulations of dwarf and Milky Way-mass galaxies}}, \href{https://doi.org/10.48550/arXiv.2401.08750}{\emph{arXiv e-prints} (2024) arXiv:2401.08750} [\href{https://arxiv.org/abs/2401.08750}{{\ttfamily 2401.08750}}].

\bibitem{astropy:2018}
{Astropy Collaboration}, A.M.~{Price-Whelan}, B.M.~{Sip{\H{o}}cz}, H.M.~{G{\"u}nther}, P.L.~{Lim}, S.M.~{Crawford} et~al., \emph{{The Astropy Project: Building an Open-science Project and Status of the v2.0 Core Package}}, \href{https://doi.org/10.3847/1538-3881/aabc4f}{\emph{\aj} {\bfseries 156} (2018) 123} [\href{https://arxiv.org/abs/1801.02634}{{\ttfamily 1801.02634}}].

\bibitem{reback2020pandas}
{The pandas development team}, \emph{pandas-dev/pandas: Pandas},  Feb., 2020.
\newblock 10.5281/zenodo.3509134.

\bibitem{harris2020array}
C.R.~Harris, K.J.~Millman, S.J.~van~der Walt, R.~Gommers, P.~Virtanen, D.~Cournapeau et~al., \emph{Array programming with {NumPy}}, \href{https://doi.org/10.1038/s41586-020-2649-2}{\emph{Nature} {\bfseries 585} (2020) 357}.

\bibitem{2020SciPy-NMeth}
P.~Virtanen, R.~Gommers, T.E.~Oliphant, M.~Haberland, T.~Reddy, D.~Cournapeau et~al., \emph{{{SciPy} 1.0: Fundamental Algorithms for Scientific Computing in Python}}, \href{https://doi.org/10.1038/s41592-019-0686-2}{\emph{Nature Methods} {\bfseries 17} (2020) 261}.

\bibitem{Hunter:2007}
J.D.~Hunter, \emph{Matplotlib: A 2d graphics environment}, \href{https://doi.org/10.1109/MCSE.2007.55}{\emph{Computing in Science \& Engineering} {\bfseries 9} (2007) 90}.

\bibitem{raybaut2009spyder}
P.~Raybaut, \emph{Spyder-documentation}, {\emph{Available online at: pythonhosted. org} (2009) }.

\end{thebibliography}\endgroup

\end{document}